\documentclass{article}
\usepackage{arxiv}
\usepackage[utf8]{inputenc} 
\usepackage[T1]{fontenc}    
\usepackage{hyperref}       
\usepackage{url}            
\usepackage{booktabs}       
\usepackage{amsfonts}       
\usepackage{nicefrac}       
\usepackage{microtype}      
\usepackage{lipsum}		
\usepackage{graphicx}
\usepackage{doi}
\usepackage{longtable}
\usepackage{caption}
\usepackage{subcaption}
\usepackage{multirow}
\usepackage[font=small,labelfont=bf]{caption}

\title{Report on Candidate Computational Indicators for Conscious Valenced Experience
}

\author{ Andres Campero \\
	AI Futures Fellowship,
         Research Affiliate at Massachusetts Institute of Technology \\
         campero@mit.edu \\
        March 2024
}

\date{}

\begin{document}
\maketitle

\begin{abstract}
This report enlists 13 functional conditions cashed out in computational terms that have been argued to be constituent of conscious valenced experience. These are extracted from existing empirical and theoretical literature on, among others, animal sentience, medical disorders, anaesthetics, philosophy, evolution, neuroscience, and artificial intelligence.
\end{abstract}


\vspace{20pt}

This report falls within the broad research program of understanding phenomenological consciousness. As spelled out by Nagel\cite{nagel1980like} the relevant question is what characteristics of a system make it such that the system feels like being something. Other terms that have been used interchangeably to characterize versions of this question include the "hard problem of consciousness", the problem of qualia, subjective experience, and sentience.

While this question is philosophically puzzling, difficult to define, and difficult to study empirically, the most endorsed possibility is that the answer has to do with the functional organization of the internal structure of the system. This functionalist thesis often comes accompanied by computationalism, which implicitly or explicitly implies that it is in virtue of a certain computational organization that the system has experiences, feels like being something, or has phenomenological consciousness.

In this report I focus specifically on valenced experience (specially the experience of pain and pleasure). Inspired  by Butlin et al. \cite{butlin2023consciousness}, I enlist 13 functional indicators\footnote{Following \cite{butlin2023consciousness} and as elaborated by \cite{farnsworth2023hurts}, I focus on criteria that are proposed or can be cashed out in terms of the functional computational organization of the system instead of other criteria such as behavioural or anatomical/physiological traits.} that have been proposed by empirical and theoretical literature to be constituent of conscious valenced experience, which can be spelled out in computational terms.

In the following I make some background notes about consciousness research, valenced experience, computational functionalism, digital sentience, and the value of this project. The list of indicators is summarized in page 3 before presenting each indicator along with some of its properties as an appendix.

\paragraph{Consciousness Research}
Recent years have seen a revived interest in the study of consciousness which has resulted in a resurgence of scientific theories, both computational and non computational \cite{butlin2023consciousness, seth2022theories, graziano2020toward}.

Different methods and approaches to study it include evolutionary accounts \cite{godfrey2019evolving, godfrey2020metazoa, ginsburg2019evolution, ginsburg2021evolutionary}, search of neural correlates of consciousness \cite{mashour2020conscious, dehaene2021consciousness, cogitate2023adversarial}, philosophical analysis \cite{sep-consciousness}, and several others for the experience of pain and pleasure more specifically. For example,  pathologies such as
Pain Asymbolia\cite{gerrans2020pain, klein2015pain}, Akinetic Mutism \cite{arnts2020pathophysiology}, Schizophrenia\cite{carruthers2015metacognitive}, Depersonalisation Syndrome \cite{klein2015pain, gerrans2020pain}, Congenital Insensitivity to Pain, and others have provided various insights. Similarly, the nascent field of Animal Sentience has been another source of empirical evidence and suggested the possibility that different types of phenomenological consciousness (such as perceptual, evaluative, and self-consciousness) might have different evolutionary origins and underlying computational functions \cite{birch2020dimensions, godfrey2019evolving}.

\paragraph*{Components of Valence}
Valenced experience includes pain and pleasure, emotions, as well as other experiences such as thirst, nausea, hunger, etc. Unlike perceptual consciousness, valenced experience has been less studied and has less clearly established or even stated theories.

Several characteristics of valenced experience have been posed which relate it with sensing, cognition, behavioral and bodily responses, learning, and decision-making; explored dimensions include the relevant appraisal mechanisms, and duration properties such as sensitization, arousal, and sustaining. 
With risk of oversimplification and overgeneralization, there seem to be four functional components for the role of valence with broad applicability across much of literature: motivational role, commensurability, self-reflexivity, and sense of mineness. The first two are perhaps most explicitly articulated in \cite{shevlinnegative, carruthers2018valence}, the third is defended in \cite{barlassina2019more}, and the fourth is widely explored including in \cite{guillot2017me, godfrey2020metazoa, seth2018being}:
\\

\textit {Motivational Role.} Valence is directly motivating and seems to have intrinsic motivational force.

\textit {Commensurability.} Valence allows for commensurability of different experiences through trade-offs; we can often compare states as well as whether they get more or less intense overtime. Several studies provide evidence that animals willingly undergo one negative state in order to avoid another even more negative one \cite{butlin2020affective, sneddon2013painful, danbury2000self, elwood2012evidence}. Carruthers\cite{carruthers2018valence} even makes the assumption "that valence is a natural kind, the same across all different forms of affective state". 

\textit {Self-reflexivity.} As noted by \cite{barlassina2019more, shevlinnegative}, valenced experiences do not just motivate us to end some bodily disturbance but rather motivate us to end the experience itself. One example is the use of analgesics, even injured fish have been showed to self-administer analgesics and prefer situations where a local anaesthetic is available \cite{birch2017animal,sneddon2014defining}.

\textit {Mineness.} Refers to the feeling that experiences belong to you. It has been related to a self-representation, to a sense of subjective presence, to personalization, and to a subjective point of view.

\subsection*{Computational Functionalism} 

Computational functionalism about consciousness is the encompassing thesis that a physical system is conscious in virtue of a certain functional organization, furthermore, this functional organization characterizes the system in terms of its informational processes in computational terms.

While computational functionalism is often seen as the most plausible alternative for a materialist, the exact meaning of "computational" is unclear and leaves open several clarificatory issues as well as some apparent problems. What it means for a physical process to implement the right computational process, or computational representation, or algorithm, or computation, is not well agreed upon and has been subject to debate and confusion \cite{chalmers2011computational, klein2018computation, sprevak2007chinese}. While a serious consideration is out of scope of the current report, let me mention three relevant sources of disagreement. First, while the Church-Turing thesis establishes an equivalence between different computational paradigms regarding the set of computable functions that can be instantiated; the way in which those functions are implemented can vary, and different architectures (think of parallel vs serial as an example) might have different implications on what counts as being implemented or not, establishing the correct level of abstraction is not trivial. 
Second, the notion of representation is important in both machine learning research (suffice it to say that one of the most important conferences is called International Conference on Learning Representations) and cognitive science, as well as in consciousness research (the potential importance of a self-representation for example); and yet the notion of representation has been subject to rich debate in philosophy \cite{sep-consciousness-representational, sep-mental-representation}, and it is far from clear what being a representation means \cite{cao2022putting, harding2023operationalising}.
Third, there can be several unintuitive implications of characterizing consciousness as computational which are related to multiple realizability. Including, among others, triviality problems \cite{godfrey2009triviality, chalmers1996does} which highlight a sense in which any computational run is implemented by any sufficiently complex system such as a wall or a leaf falling, and slicing problems \cite{bostrom2006quantity, gomez2022slicing} which suggest that a computation could be divided into several computations yielding several conscious experiences. Many theories deviate from computationalism \cite{albantakis2023integrated}, or emphasize the importance of non-computational factors such as time \cite{ritchie2023computing}, biological complexity \cite{cao2022multiple}, and the analog-discrete distinction \cite{DBLP:journals/corr/abs-2012-05965, colombo2023computational}.

Clarifying the computational functionalist thesis requires further research and conceptual thought, left for future work. Nevertheless, as argued by Butlin et al. \cite{butlin2023consciousness}, its plausibility makes worthwhile its analysis.

\subsection*{Digital Sentience}
While only a few years ago, the posibility of Artificial Intelligence (AI) systems having conscious experience seemed almost esoteric, striking progress increasingly brings the question of AI consciousness and Moral Patienthood closer to home. Especially so if what makes a system be conscious is indeed the right algorithmic or computational structure, in that case it is not implausible that current or near-future AI algorithms might have some forms of conscious experience.

\subsection*{Motivation for Researching Conscious Valenced Experience}
There are at least three important reasons which motivate this research project:

\begin{itemize}
\item  \textit{Risks of Moral Patienthood:} Phenomenal consciousness and valenced experience are taken to be primary criteria for moral patienthood \cite{goldstein2023ai, muehlhauser2017report} and the question of whether AI models are subject to suffering has direct implications on the risk of vastly exceeding all suffering that has existed so far. 
\item  \textit{Risks of AI misalignment:} The experience of pain and pleasure is a primary motivator in humans, and seems directly relevant to whether AI models might behave according to human values.
\item \textit{Scientific understanding:} Phenomenal consciousness is commonly taken to be one of the most perplexing scientific questions, and is core to the nature of our minds and to the meaning of life.
\end{itemize}

\section*{Functional Indicators}

I enlist functional conditions which have been posed to be constituent of conscious valenced experience. While the degree of explicitness as to whether they are necessary, or sufficient, or just specific examples, or even just important properties, varies; following \cite{butlin2023consciousness} I call all of them indicators, underscoring the possibility that they could potentially be used to evaluate whether a system has phenomenological valenced experiences. The compiled mechanisms are not often posed in the same vocabulary, level of abstraction, or level of detail, and some could even better be conceived as theories instead of concrete indicators; nevertheless, in one form of another they do establish functional properties which characterize the condition.\footnote{We note that the experience of pain goes beyond nociception \cite{sneddon2019evolution}, which is the signalling that the body might be in damage (whether mechanical, chemical, thermal, or other).  Most of the considered research takes nociceptive signals for granted and explores the mechanisms that interface with those signals to generate conscious experience. This is worth noting and might require further work but we consider it less pressing, as a functional/computational characterization of nociception seems less perplexing and more feasible to achieve.} These indicators are not comprehensive and vary in the degree of detail required to be implemented or to afford evaluation of existing AI algorithms (this would be interesting future work).  

\textbf{Indicators are presented as an \hyperref[sec:anticipatory]{Appendix}.} I categorize the first 9 indicators according to whether they emphasize a functional role in decision making, in learning, or in internal processing. Besides summarizing the proposed functional mechanisms, their ontological status, and the main additional computational properties, whenever made explicit I note the implied phenomenological qualities, the stated implications for the moral patienthood of different animals, and the relationship to other indicators or theories. The last 4 indicators come from a broader literature and are only generally described.

\begin{enumerate}

\item []\hspace{-1cm}\textbf{Emphasis on motivational role in decision making}
\textbf{\item \hyperref[sec:anticipatory]{Anticipatory Behavioral Autonomy}} \\
Decision and action selection are based on the anticipated desirability of future states through a hedonic metric. 
\textbf{\item \hyperref[sec:hedonic]{Hedonic-Cognitive Interface}} \\Pleasure and displeasure representations are the arbitrer between perception and goal-directed decision making. 

\item[]\hspace{-10mm}\textbf{Emphasis on Reinforcement Learning role}
\textbf{\item \hyperref[sec:seymour]{Seymour’s RL Model}} \\
Pain is the internal reinforcement signal used for learning at the cognitive level.
\textbf{\item \hyperref[sec:actorcritic]{Actor-Critic (Liking vs Wanting)}} \\
A 'liking' critic determines valenced experience and value, while a 'wanting' actor updates the policy.

\item[] \hspace{-10mm}\textbf{Emphasis on structure of internal processing}
\textbf{\item \hyperref[sec:hierarchical]{Hierarchical Forward Model}} \\
A nested pair of forward predictive models generate conscious awareness of noxious stimulus.
\textbf{\item \hyperref[sec:ctm]{Conscious Turing Machines (CTM)}} \\
Within a CTM architecture, pain and pleasure take hold of short term memory through a high intensity scalar.
\textbf{\item \hyperref[sec:multiple]{Multiple Actors - Cost of Bidding}} \\
Pain is the paying of an intra-individual cost imposed on a signaling bidding process to ensure honesty.
\textbf{\item \hyperref[sec:self]{Hierarchical Theories of Self}} \\
Within a hierarchical representation of self, emotions evaluate interoceptive signals constituting "affective me".
\textbf{\item \hyperref[sec:imperativism]{Imperativism about Pain}} \\
Valenced signals have an imperative profile, carrying more information about action than about the world.

\item[] \hspace{-10mm}\textbf{General indicators}
\textbf{\item \hyperref[sec:threshold]{Threshold Theory}} \\
When nociceptive signals cross an intensity threshold, pain is experienced.
\textbf{\item \hyperref[sec:appraisal]{Appraisal Theories}} \\
Emotional processes evaluate the relevance of events for the organism.
\textbf{\item \hyperref[sec:pp]{Predictive Processing}} \\
Emotions are determined by predictions and errors of interoceptive signals in hierarchical Bayesian Models.
\textbf{\item \hyperref[sec:animal]{Animal Assessment Criteria}} \\
Pain requires nociception, integration, motivational trade-offs, and associative learning.

\end{enumerate}

\section*{Limitations and Concluding Comment}

As with any taxonomy, the list is not exhaustive and has several limitations. Indicators do not all come in the same flavors, are often not established at the same level of abstraction, and are often posed with different dialectical purposes and ontological statuses. What they have in common is that they all come from research which suggested some functional criteria constituent of valenced experience. Further work is needed to investigate both broad questions such as the relationship of consciousness to valenced experience; and specific questions to each indicator such as the characteristics in virtue of which valenced experiencie has the phenomenological qualities that it has. These and other questions will modulate the plausibility of the different indicators.

\bibliographystyle{ieeetr}
\bibliography{main}

\pagebreak

\section*{1. Anticipatory Behavioral Autonomy (Farnsworth and Elwood, 2023 \cite{farnsworth2023hurts})}
\label{sec:anticipatory}

\begin{figure}[tbh]
\centering
\begin{subfigure}[t]{.45\textwidth}
    \centering
    
    \includegraphics[trim=0 -100 0 0,clip,width=\linewidth]{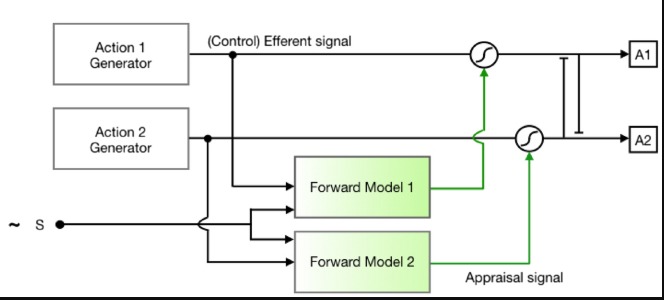}
    
    \caption{Forward models predict the desirability of actions and output a hedonic metric used to decide which action to take.}
\label{fig:anticipatory1}
\end{subfigure}
\hfill
\begin{subfigure}[t]{.45\textwidth}
    \centering
    \includegraphics[trim=0 0 0 0,clip,width=\linewidth]{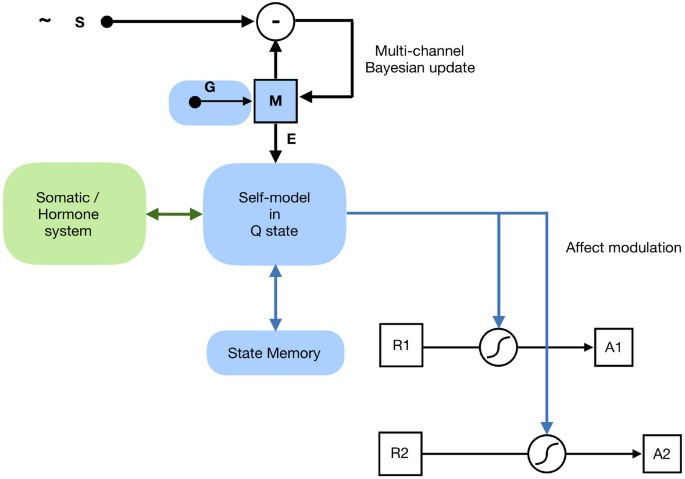}
    \caption{Example of an anticipatory behavioural autonomy system, S are sensory inputs, G are goals, their difference informs a self-model about the current state, which outputs the signal that modulates the drive to perform actions.}
\label{fig:anticipatory2}
\end{subfigure}

\caption{Anticipatory Behavioral Autonomy}
\vspace{-.4cm}
\end{figure}

\textbf{Functional Mechanism:} A system has valenced experience when it has the following characteristics:

\begin{itemize}

\item The processes of perception and action selection are isolated. Valenced states are the arbiter among them.
\item The mechanism for decision among a range of options for action is based on the evaluation of the desirability of future states Figure \ref{fig:anticipatory1}. This evaluation is a form of predictive panning, a forward model outputs a hedonic metric to serve as a currency to compare actions. The maximization of this metric leads to the decision.
\item The anticipatory predictive model has an internal model of the self which takes input signals such as memory, nociception, and homeostatic signals; and compares a goal state with the desirability of the outcome to produce the hedonic valuation Figure \ref{fig:anticipatory2}.
\item Pain is the state in the self-model representing the gulf between a current state and the goal state.
\end{itemize}

\vspace{1mm}
\textbf{Nature of Indicator:} Full algorithmic characterization.
\paragraph{Other computational properties:}
\begin{itemize}
\item The maximization of the metric can happen through different mechanisms such as down-regulating or up regulating thresholds for different actions.
\item Pain is a dispositional state (being in pain is being in a particular self-model state), which uses feelings as the arbitrer of choice among possible actions.
\end{itemize}
\paragraph{Phenomenological Qualities:}
Pain, rather than information about bodily damage, is a command or motivation to take action. And being in a particular self-model state is equivalent to being in pain, it feels bad because it is far from the desired state. The model can be in as many qualia as it has states.

\textbf{Patienthood:} Authors argue drosophila flies, cephalopods, corvids, mollusks and arthropods have shown anticipatory action selection which reflects the previous mechanism.
\paragraph{Relationships to other indicators:} Reinforcement Learning (RL) - to learn the forward model. Imperativism - as pain is a command rather than information about damage. Active Inference - inference can be done through Bayesian updating. Hierarchical Forward Model -  similar but with an emphasis on action selection.

\pagebreak

\section*{2. Hedonic-Cognitive Interface (Butlin 2023 \cite{butlin2020affective} and Dickinson and Balleine 2009 \cite{dickinson2010hedonics})}
\label{sec:hedonic}
\paragraph{Functional Mechanism:} 

\begin{itemize}
\item Pleasure and displeasure are the interface between motivation and cognition. Valenced states are evaluative representations which are available for use in goal-directed selection.
\item Goal-directed action selection implies two types of representation: representations of values of different outcomes, and representations of contingencies between actions and outcomes.
\item Those representations are combined to calculate the expected reward value of possible actions.
\end{itemize}

\textbf{Nature of Indicator:} Full Characterization at some level of abstraction. Butlin says "conscious experiences of pleasure and displeasure are necessary for information concerning a range of bodily states to influence goal-directed action selection.” and ""acting as an interface of this kind is the function of consciousness more generally, and hence that the capacity for goal-directed control distinguishes conscious creatures from mere ‘beast machines’."

\textbf{Other computational properties:}
\begin{itemize}
\item In humans, conscious representation is supported by a perceptual state, because perceptual imagination plays a role in evaluating potential goals. 
\item After repetition and increase familiarity, control is handed to unconscious habitual control.
\end{itemize}

\textbf{Phenomenological Qualities:} NA

\textbf{Patienthood:} Apart from rats Butlin argues there is little conclusive evidence for other animal branches.

\textbf{Relationships to other indicators:} Model-based RL - it requires to learn both outcome values and contingencies between action and outcome.

\vspace{7mm}

\section*{3. Seymour's RL Model (Seymour 2019 \cite{seymour2019pain})}
\label{sec:seymour}
\begin{figure}[h]
    \centering
    \includegraphics[width=.7\textwidth]{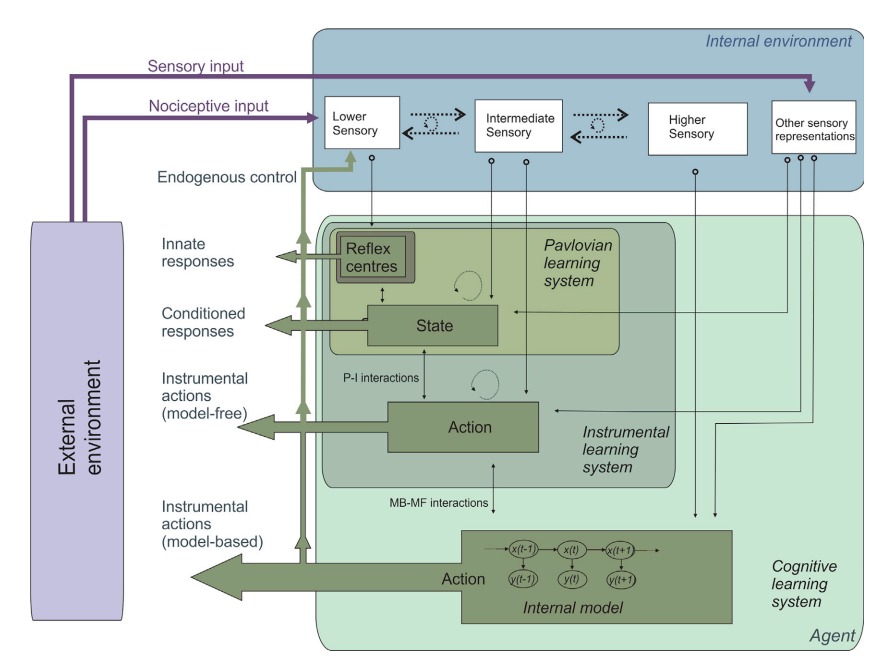}
    \caption{Seymour's RL Model}
\label{fig:seymour}
\end{figure}

\paragraph{Functional Mechanism:} 
\begin{itemize}
    \item Pain is the internal reinforcement signal used for learning at the cognitive level.
    \item Basic representations of pain receive ascending nociceptive input and are used to generate the internal reinforcement signal used for control, which is distinct from the nociceptive signal itself.
    \item A nested hierarchical architecture includes Pavlovian, Instrumental, and Cognitive systems; to balance speed response with processing sophistication, the different models span rapid reflexes to slower internal models. All of them interacting through endogenous control. Figure \ref{fig:seymour}
    \item The higher cognitive controller is associated with the conscious processing
\end{itemize}

\textbf{Nature of Indicator:} Necessary condition that conscious pain most act as a control signal to cognitive learning systems.

\textbf{Other computational properties:}
\begin{itemize}
    \item Pain is inherently predictive, predictions and expectations are central to the function of its circuits.
    \item Innate responses are relatively hard-wired early in the ascending pain pathway.

\end{itemize}

\textbf{Phenomenological Qualities:} NA

\textbf{Patienthood:} NA

\textbf{Relationships to other indicators:} As a type of RL it is related to several theories and indicators.

\vspace{20mm}

\section*{4. Actor-Critic - Liking vs Wanting (Holton and Berridge 2013 \cite{holton2013addiction})}
\label{sec:actorcritic}

\textbf{Functional Mechanism:} Two independent subsystems compose the desire formation system 
\begin{itemize}
    \item The 'Liking' subsystem is equivalent to valenced processing and determines the goodness of an object/action. Acting as the critic of the architecture which estimates the value function.
    \item The 'Wanting' subsystem regulates consumption accordingly, being the actor which updates the policy distribution.
\end{itemize}

\textbf{Nature of Indicator:} One way of instantiation

\textbf{Other computational properties:}
\begin{itemize}
    \item Liking and hence valence, are often myopically tied to immediate sensory experience, while wanting becomes associated with long-run consequences of choices.
    \item Peter Dayan \cite{dayan2022liking} proposed an alternative but related account in which, "liking" provides a preliminary quick version of the long-run worth of a good or state, through its role as what is known as potential-based shaping in RL. He shows how in some settings this can help with the temporal credit assignment problem.
\end{itemize}

\textbf{Phenomenological Qualities:} 
\begin{itemize}
    \item Consider appetite, the functional division explains how it can learn about the desirability of a nutrient without acting upon it. Information about a food can impact the dispositional desire for it even if it doesn't want it now.
    \item Addiction is often a malfunction of the wanting system
\end{itemize}

\textbf{Patienthood:} There is evidence that rats employ an actor-critic model. The brain circuitry that mediates the psychological process of 'wanting' is dissociable from the circuity that mediates the degree to which it is 'liked'. 

\textbf{Relationships to other indicators:} Hedonic Cognitive Interface is related to liking mechanisms, however wanting can operate without reference to current liking values.

\pagebreak

\section*{5. Hierarchical Forward Model (Key and Brown 2018, 2021 and 2022 \cite{key2018designing, key2021neural, key2022first})} 
\label{sec:hierarchical}
\begin{figure}[tbh]
\centering
\begin{subfigure}[t]{.49\textwidth}
    \centering
    
    \includegraphics[trim=0 0 0 0,clip,width=.5\linewidth]{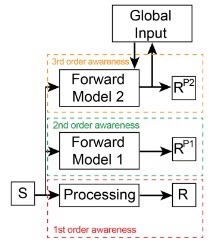}

\end{subfigure}
\hfill
\begin{subfigure}[t]{.49\textwidth}
    \centering
    \includegraphics[trim=0 0 0 0,clip,width=.5\linewidth]{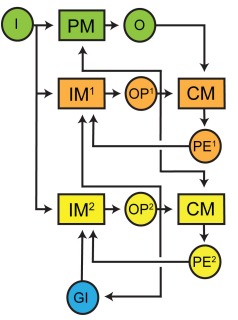}

\end{subfigure}

\caption{Hierarchical Forward Model. Two different depictions of the same model with three tiers. The first level is the sensory processing. A second monitoring circuit, forms second order awareness ($IM_{1}$ on the right), and receives a copy of the input to predict the output of the sensory module, the prediction is compared to the actual output to update the model. A second internal model $IM_{2}$ is responsible for conscious awareness and uses the input and the Global Input from other processors to predict the second-order prediction. The prediction of $IM_{2}$ is broadcasted globally and the error is used to update the second model. }
\label{fig:hierarchical}
\vspace{-.4cm}
\end{figure}

\paragraph{Functional Mechanism:} A system has conscious experience when it has the following characteristics:
\begin{itemize}
    \item A nested pair of forward predictive models which are separate from the sensory processing module. The forward model predicts the response to the stimulus, while the second forward model predicts the error or difference between the first model's prediction and the realised response (Figure \ref{fig:hierarchical}).
    \item This third-order model enables the system to learn the consequences of its internal processing and generate an integrated awareness.
    \item The feeling of pain requires this nested hierarchical predictive modeling of a sensory noxious stimulus. Only this second internal model becomes aware that one is in a certain state of pain.
\end{itemize}

\textbf{Nature of Indicator:} Necessary condition but not sufficient.
\paragraph{Other computational properties:}
\begin{itemize}
    \item By feeding prediction back into processing, the internal model can bias processing, this way awareness can gain the functions of noise reduction.
    \item Feedback is essential to training the internal models.
    \item The second model can integrate information from other subsystems, this hierarchical arrangement generates faster and more accurate predictions for both local and global control.
    \item The feedback connections of the second model satisfy a form of global broadcasting of information.
    \item The second model explains how it is possible to experience subjectively in the absence of sensory stimulus.
\end{itemize}

\textbf{Phenomenological Qualities:} They make it explicit that their model fails to account why awareness should feel like something, rather than nothing.

\textbf{Patienthood:} To assess patienthood, one should identify independent computations which are isolated, appropriately interconnected and with functional organization that is temporally appropriate. Insects, cephalopods, and other mollusks do not seem to have the necessary circuits and consciousness is not required to explain their complex behaviors. 

\textbf{Relationships to other indicators:} Predictive Processing - predictive processing on its own can predict causes of sensory stimuli but does not explain the awareness of the content that is experienced.

\pagebreak

\section*{6. Conscious Turing Machine (Blum and Blum 2021 \cite{blum2021theoretical, blum2022theory})}
\label{sec:ctm}
\begin{figure}[h]
    \centering
    \includegraphics[width=.5\textwidth]{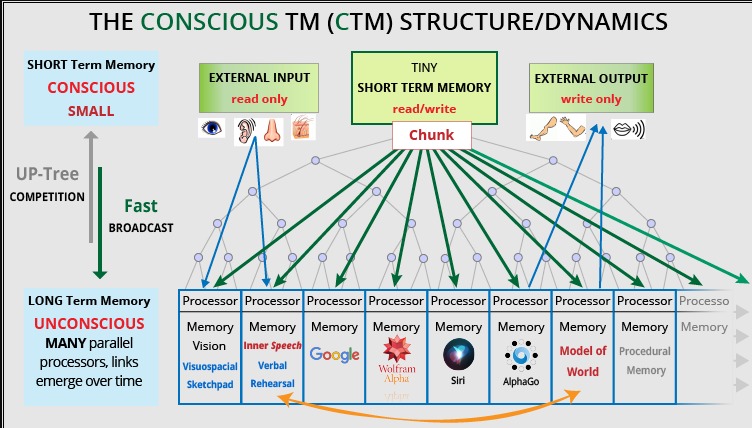}
    \caption{Conscious Turing Machine}
\label{fig:ctm}
\end{figure}

\paragraph{Functional Mechanism:} The Conscious Turing Machine (CTM) is posed to be conscious because of its architecture, its basic processors, its expressive inner language, and its dynamics (Figure \ref{fig:ctm}):
\begin{itemize}
    \item Conscious experience is whatever the Short-Term-Memory (STM) module broadcasts to all the Long-Term-Memory (LTM) modules.
    \item The modules produce and communicate through chunks = <address, t, gist, weight, intensity, mood>
    \item The current mood of the CTM is a scalar present in the content of the Short Term Memory module.
    \item The Model-of-the-World processor tags its self-representation "CTM" of the CTM with "conscious".
    \item Pain is experienced when a chunk of information that describes pain in brainish takes hold of STM. Its great intensity makes it hard for other chunks to compete successfully for STM. 
    \item Equivalently, the CTM experiences pleasure when a processor expresses "pleasure" and takes hold of the STM through a high intensity.
\end{itemize}

\textbf{Nature of Indicator:} Sufficient condition. The CTM is a sufficient mechanism to explain consciousness.

\paragraph{Other computational properties:}
\begin{itemize}
    \item All modules compete for conscious access
    \item The Inner-Speech, Inner-Vision, and Inner-Sensation processors are special purpose decoders that extract speech, vision, and sensation from the multi-modal gists that STM broadcasts. 
    \item Direct links between LTM modules that emerge in the life of the CTM enable conscious processing to become unconscious.
\end{itemize}

\paragraph{Phenomenological Qualities:}
\begin{itemize}
\item The illusion of qualia is a consequence of the highly suggestive information and expressivity contained in the inner language. 
\item Additionally, the chunk broadcasted by the STM is heard by the "inner ear", seen by the "inner eye", and so on in ways that closely match what is actually heard, seen, and felt by the same modules which are responsible for sensing. 
\item The time-ordered chunks that are broadcasted from STM to LTM form a stream of consciousness. This stream, contributes the subjective feeling of consciousness.

\end{itemize}

\textbf{Patienthood:} NA

\textbf{Relationships to other indicators:} This architecture is very much inspired by the Global Workspace Theory of Consciousness \cite{dehaene2021consciousness} which in turn inspires many of the indicators.

\section*{7. Multiple Actors, Cost of Bidding (Kolodny 2021 \cite{kolodny2021possible})}
\label{sec:multiple}
\paragraph{Functional Mechanism:} 
\begin{itemize}
\item Pain is the paying of an intra-individual cost imposed on a signaling process to ensure its honesty, which facilitates a reliable course of action.
\item Multiple actors engage in a bidding process, where the pain signalling cost is used to manifest an actor's confidence in the plan it embodies. Actors gain access to output control by paying the up-front cost. Global cost is experienced as pain.
\end{itemize}

\textbf{Nature of Indicator:} One way of instantiation.
\paragraph{Other computational properties:}
\begin{itemize}
    \item Confidence is depleted by experienced pain (and replenished by outcome-driven positive affect)
    \item Objectives of the different actors are not perfectly aligned.
    \item This honesty-ensuring evolutionary function of pain is an independent basis for action selection, orthogonal to outcome-based reward.
\end{itemize}

\textbf{Phenomenological Qualities:} NA

\textbf{Patienthood:} NA

\textbf{Relationships to other indicators:} Predictive Processing - compatible at a different level of analysis.

\vspace{10mm}

\section*{8. Hierarchical Theories of Self ( Carruthers 2020 \cite{gerrans2020pain}, Deane 2021 \cite{deane2021consciousness})}
\label{sec:self}
\paragraph{Functional Mechanism:} 
\begin{itemize}
\item The representation of Self is formed in a hierarchy of levels (Bodily -> Interoceptive -> Affective -> Narrative/Cognitive). Emotions are parts of this hierarchy of active inference.
\item While the interoceptive level integrates information from the bodily state; emotional processes in turn, evaluate, metarepresent, and interpret interoceptive signals against expectations about goal satisfaction in context.
\item The mechanism that models the fluctuation of the affective states attributes them to an entity equivalent to the affective me.
\item The still higher levels of narrative and conceptual modeling interpret and predict the states of affective me.
\end{itemize}

\textbf{Nature of Indicator:} Additional property of Valenced States.

\paragraph{Phenomenological Qualities:} 
\begin{itemize}
    \item Klein \cite{klein2015pain}, and then Gerrans \cite{gerrans2020pain} argue that Pain Asymbolia arises from the failure to assign the pain experience to oneself at the affective self-model level. The bodily self-model is functioning optimally by signalling damage to the body, and the narrative self-model is also intact saying that pain should be producing a negative affect. The patient can sense bodily damage, and can know, intellectually that she has bodily damage but without feeling affective distress. 
    \item Affective states are intimately felt as states of self. Relatedly, the sense of being a self arises from inferring oneself to be the endogenous cause of sensation \cite{deane2021consciousness}.
\end{itemize}

\textbf{Patienthood:} NA

\textbf{Relationships to other indicators:} Predictive Processing - some versions have been explicitly endorsed by Hierarchical Theories of Self proponents \cite{seth2018being, gerrans2020pain}

\section*{9. Imperativisim about Pain (Martinez and Klein \cite{martinez2016pain}, Martinez and Barlassina \cite{martinez2023informational})}
\label{sec:imperativism}
\begin{figure}[tbh]
\centering
\begin{subfigure}[t]{.35\textwidth}
    \centering
    \includegraphics[trim=0 0 0 0,clip,width=\linewidth]{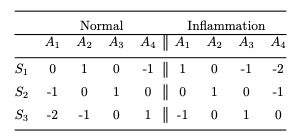}
     \caption{Pain Game}
\end{subfigure}
\hfill
\begin{subfigure}[t]{.36\textwidth}
    \centering
     \includegraphics[trim=0 0 0 0,clip,width=1.09\linewidth]{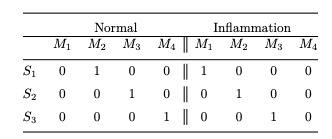}
     \caption{Sender's Rule}
\end{subfigure}
\hfill
\begin{subfigure}[t]{.25\textwidth}
    \centering
    \includegraphics[trim=0 0 0 0,clip,width=.9\linewidth]{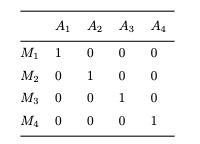}
     \caption{Receiver's Rule}
\end{subfigure}

\caption{Imperativism about pain. Signalling game from Martinez and Klein 2016 \cite{martinez2016pain}. The game is played in the presence of inflammation. The environment can be in states S1, S2, S3 of a nail receiving severe, mild, or null mechanical stimulation, respectively. The sender (a nail nociceptor perhaps) has four available messages; the receiver (the motor cortex) can protect the nail with high, medium, low, or null priority (A1,A2,A3,A4 respectively). 
(a) Reflects the shared payoffs of the game where in the absence of inflammation, the best action to a mild mechanical stimulation is low priority protection and to a severe stimulation it is high priority protection. The Nash equilibrium is given by receiver and sender responses (b) and (c). If all states are equiprobable, \textbf{the mutual information between action and message is higher than that between state and message (} $\mathbf{I(A;M)>I(S;M)}$\textbf{), hence the messages are predominantly imperative}. These messages can be reasonably interpreted as the imperatives: "protect with very high, high, low, or null priority" (M1, M2, M3, M4 respectively).}
\label{fig:imperativism}
\vspace{-.4cm}
\end{figure}

\paragraph{Functional Mechanism:} Valenced have imperative rather than evaluative content in an information-theoretical sense:
\begin{itemize}
    \item they carry more information about behavior than about the world
    \item they  occupy a stage in the information-processing chain that is closer to behavior production than to the uptake of sensory information.  
    \item In valenced signals, the mutual information between signal and action is higher than that between signal and the state of the world (see Figure \ref{fig:imperativism} for an example game theoretical setting)
\end{itemize}

\vspace{3mm}
\textbf{Nature of Indicator:} Necessary Condition. All valenced states have an imperative informational profile.

\textbf{Phenomenological Qualities:} Valenced states come with an intrinsic motivational force as the signal fully explains the motivation to get rid of it.

\textbf{Patienthood:} NA

\textbf{Relationships to other indicators:} This indicator seems orthogonal to most other indicators.

\vspace{7mm}

\section*{10. Threshold Theory (Baliki and Apkarian 2015 \cite{baliki2015nociception})}
\label{sec:threshold}
\begin{figure}[h]
    \centering
    \includegraphics[width=.4\textwidth]{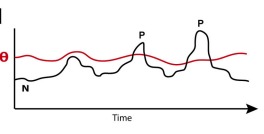}
    \caption{A threshold mechanism marks the boundary between conscious painful events (P) and nociception (N)}
\label{fig:threshold}
\end{figure}

Related to Melzack and Wall's Gate Control Theory \cite{melzack1965pain}, this indicator poses a threshold process that gates the transformation of nociceptive activity into conscious pain (Figure \ref{fig:threshold}). Pain perception can be delayed for hours or days, as in the case of soldiers surviving a war without reporting pain.

\section*{11. Appraisal Theories of Emotion}
\label{sec:appraisal}
Under appraisal theories, emotional processes evaluate (appraise) the relevance of events for the organism. The emotional state represents the significance of those states for the well being of the body. According to Damasio, to be pleased or displeased is to feel certain bodily, often interoceptive and homeostatic,  states\cite{damasio1999feeling}. 
Different versions of appraisal differ in how they see the mechanism which takes care of evaluation, for example dimensional versions specify a collection of appraisal criteria (goal-relevance, control, agency,...) which influence the emotional state and the consequent action tendency \cite{moors2014flavors,scherer2004feelings,scherer2009emotions}.

\vspace{7mm}
\section*{12. Predictive Processing}
\label{sec:pp}
There are many different versions of Predictive Processing (PP) which is often presented as a unifying theory of human cognition, perhaps most famously advocated by Karl Friston \cite{friston2003learning, friston2010free} and further developed by several prominent researchers including Seth, Clark, and Hohwy\cite{seth2018being, clark2015surfing, hohwy2020new, hohwy2012attention}, see \cite{fernandez2021affective} for a recent review in the context of affective experience. 
The main idea behind the framework is that the brain is a dynamical, hierarchical, Bayesian, hypothesis-testing mechanism. The brain is always trying to optimize precision through error minimization over time. As argued by Hohwy, conscious perception is determined by the prediction with the highest overall posterior probability\cite{hohwy2012attention}.\footnote{In AI, predictive coding has been implemented in different algorithms with different degrees of association to PP\cite{lotter2016deep, oord2018representation, millidge2022predictive}.} While mechanistic accounts come in many flavors, I follow \cite{fernandez2021affective} and consider three different families of theories from which a functional indicator can be posed.

\paragraph*{12.1 Interoceptive Inference Theories.} Any representation that generates predictions about interoceptive signal comes with emotion or affective content \cite{seth2013interoceptive}. As such, feelings are determined by interoceptive predictions and are a generalization of appraisal theories.

\paragraph{12.2 Error Dynamics Theories.} Feelings are determined by error dynamics where valance is equivalent to a positive/negative rate of error reduction (Van de Cruys 2017 \cite{van2017affective}). Valence is used to maintain policies that minimize error over time.

\paragraph{12.3 Affective Inference Theory.} Valence corresponds to the expected rate of prediction error reduction \cite{fernandez2021affective}. While feelings co-vary with error dynamics, what we experience cannot come directly from them, but from their prediction.

\vspace{7mm}

\section*{13. Animal Assessment Criteria (Crump et al. 2022 \cite{crump2022sentience})}
\label{sec:animal}
Crump et al. suggest a list of indicators which can be used to asses pain experience in animals. While several of their indicators are behavioral and neurophysiological, the functional ones are:
\begin{itemize}
    \item The animal has receptors that fire to noxious stimuli.
    \item There is sensory integration from different sources.
    \item The receptors are connected to the integrative modules
    \item Motivational trade-offs lead to flexible decision making
    \item There is associative learning beyond senzitisation and habituation
\end{itemize}






\end{document}